\documentclass[%
 twocolumn, 
 amsmath,amssymb,
 aps, 
 prc,
floatfix,
]{revtex4-2}

\usepackage[T1]{fontenc}

\usepackage{graphicx}
\usepackage{dcolumn}
\usepackage{bm}
\usepackage{hyperref}

\usepackage{amssymb}
\usepackage{amsmath}

\usepackage{physics}
\usepackage{tabularx}
\usepackage{siunitx}

\usepackage{orcidlink}

\renewcommand{\selectlanguage}[1]{}

\begin{document}


\title{\textbf{An improved statistical treatment for radioassay-based background projection incorporating part-to-part variation} 
}%

\author{Louis Varriano\orcidlink{0000-0001-5961-0688}}
 \email{Contact author: varriano@uw.edu}
\author{Jason Detwiler\orcidlink{0000-0002-9050-4610}}%
\affiliation{%
 Center for Experimental Physics and Astrophysics, University of Washington, Seattle, WA, USA
}%

\date{\today}

\begin{abstract}
For low background experiments, prediction of the background from radioactivity in detector components is of critical importance. Radioassays are often conducted on small samples or individual parts, which may be subject to differences in production or handling resulting in a part-to-part variation. This part-to-part variation can have an impact on the total activity and is not presently captured by the averaging methods used to evaluate radioassays. We propose a new method for the averaging of radioassays to incorporate this part-to-part variation into predictions of the total activity of parts. We also  consider how to add component activities based on assays for component groups that are simulated as a single entity.
\end{abstract}

\maketitle


\section{Introduction}

Current and future low background experiments rely upon radioassays to estimate and  predict their experimental sensitivities, as well as constrain their background models; for recent examples in current and next generation dark matter and neutrinoless double-beta decay experiments, see e.g.~\cite{Leonard_2008,Abgrall:2016cct,Albert_2018,Alduino_2017,Leonard_2017,XENON_2017,Gerda_2018,Akerib_2020,Amaudruz2019,Gileva_2023,Agrawal_2024}. As experimental sensitivities have continued to improve, materials with correspondingly lower levels of radioactive impurities are required. In contemporary experiments, the required radiopurities of such materials are often close to or beyond the sensitivity of typical radioassay techniques. Furthermore, it is often possible to assay only a sample or subset of the actual components to be used in an experiment, requiring the extrapolation or scaling of the assay result to the total number of components in the experiment.

Much thought has been put into how best to use the results of assay to predict experimental sensitivities. Here, we focus on two smaller but no less important problems related to assay interpretation: 1) how to average multiple, possibly conflicting assay results of a material or part, and 2) how to combine activities of individual materials or parts into a single grouped activity for simulation or fitting purposes. Various approaches to these problems have been proposed and used in recent years, see for example~\cite{Tsang_2018,mjdstatistics}. We propose a new method for averaging assay results that incorporates part-to-part variation, which has not been previously considered to our knowledge.

In practice, the results of a particular assay can often be difficult to interpret or directly apply to an experimental projection or spectral fit. Due to the stringent radiopurity requirements, differences in handling and sample preparation can have a large impact on the results. Direct assay of as-built components is generally preferred though often impractical or impossible to perform due to the delicateness or low mass of components. There is no universally applicable recommendation for statistical treatment since considerable care and understanding must be taken in interpreting each individual assay and its relevance to the overall experiment. Nevertheless, with some basic and generic assumptions, we make some commonsense recommendations that can have broad applicability for low-background experiments. We emphasize that our primary interest is the application of assay results as inputs or constraints to simulated spectral fits and the prediction of experimental sensitivities based on simulations; a different application might reasonably come to different conclusions on the use of assay results.

For clarity, we first define the following terms. For the background model of an experiment, a {\it component group} is simulated as a single entity with activity distributed uniformly over its volume. Many component groups will be made up of {\it components}, which are different materials of construction with different activities. Component groups and components are made up of many individual identical {\it parts}.

\section{Assay results as likelihoods}\label{sec:basics}

To begin addressing the two stated problems, we must first understand how best to interpret the results of a radioassay. 
Inference on the activity of an assayed component is based on the likelihood of the assay measurement under different hypothesized activity levels. The inference made also depends on the statistical treatment employed, such as the decision whether to report a limit or a measurement, whether to use frequentist or Bayesian methods, etc., but in all cases the likelihood, based on the underlying probability distribution of the measurement itself, plays the central role.
The same likelihood, or a derivative of it such as a posterior distribution, also influences the result of background modeling fits to data as it is added as a cost function to the fit. It is therefore necessary to use a likelihood shape that approximates well the true statistical nature of the measurement, while also considering the benefit of selecting a simplified form to ease analysis and computation. Thanks to the Central Limit Theorem in probability theory, the Gaussian likelihood is very often a convenient and reasonable choice for the shape of the likelihood, so we restrict our discussion to that case here.

We focus on two of the most commonly used radioassay techniques: inductively-coupled plasma mass spectrometry (ICP-MS, e.g. \cite{Gileva_2023}) and gamma counting (e.g. \cite{Neder_Heusser_Laubenstein, Araujo_2022}). These techniques have quite different systematic and statistical uncertainties, which have been considered in the literature at length (e.g. \cite{Currie_1968, Friedlander_1981, Hurtgen_2000, Strom_MacLellan_2001, Ulianov_2015}).  
Despite this, in the authors' experience, most radioassay reports do not include the necessary details to reconstruct the likelihood function required to address the problems that are the subject of this manuscript, 
Perhaps this is because historically ``customers'' of such reported assay values simply have not requested this information, and have settled for simple limits or values with error bars. It is typically beyond their scope to re-assess the statistics of each reported assay value; it would require an extensive understanding of the details of the construction of this limit from the assays and under what conditions a limit is reported instead of a confidence interval, 
which can vary from lab to lab. Often the assays have been performed many years prior to its use in a fit or projection, and the information required to construct a detailed statistical model may no longer be available. It is preferable from this point of view to have a simple treatment that can be applied to common assay reports.

In the case of gamma assays, although performed in low-background facilities, these measurements still largely occur in a background-limited regime. Still, due to the small number of counts obtained in the most sensitive of spectrometers, a concern is the under-estimation of the activity of a sample or background. Recent recommendations \cite{Hurtgen_2000} for low statistics spectrometers advise a Bayesian treatment which applies a uniform prior on the number of counts obtained by the spectrometer, resulting in a 1$\sigma$ statistical uncertainty of $\sqrt{N+1}$ for $N$ counts obtained \cite{Friedlander_1981}. Practically, this prior only slightly inflates the statistical uncertainty obtained compared to a frequentist treatment (i.e. $\sqrt{N}$) with the largest effect at the lowest number of counts where the frequentist treatment also diverges from the large $N$ limit. More importantly, Ref. \cite{Hurtgen_2000} recommends a rather conservative treatment of upper limits when the sample count rate does not exceed the calculated level of detection due to difficulty in estimating the expected background rate, as well as the large relative fluctuation in the background rate due to its low statistics. This results in an upper limit based on a Gaussian approximation.

For ICP-MS assays, systematic uncertainties also to contribute significantly to the overall uncertainty such that these measurements are also effectively performed in a regime with Gaussian uncertainty \cite{Ulianov_2015}. Compared to gamma spectrometery, labs performing ICP-MS assays may use different practices for calculating and reporting the limit of detection \cite{Tsang_2018}. In addition, often in these assays large variation exists between different samples of the same component \cite{Christofferson:2017nih, Abgrall:2016cct} or even different replicates. We have elected to perform an averaging procedure, which would also tend toward a Gaussian likelihood in the limit of a large number of assays, as detailed in section \ref{sec:parttopart}. 

To avoid further complications, and following the recommendation of Ref. \cite{mjdstatistics}, we also suggest the reinterpretation of reported upper limits as a measurement of zero with Gaussian uncertainty. To convert from an upper limit to a corresponding 1$\sigma$ uncertainty, we use the corresponding $z$-score of the stated confidence level (CL),

\begin{equation}
    \sigma = \frac{\textrm{UL}}{\sqrt{2}\erf^{-1}(\textrm{CL})},
\end{equation}

\noindent where UL is the upper limit at confidence level CL (e.g. for 90\% confidence level, CL = 0.9). 

We illustrate this by comparing a limit obtained from a Gaussian likelihood with an identical limit obtained from a Poisson likelihood. These two likelihoods are the most important cases to consider, since both gamma assays and ICP-MS are fundamentally based on the subtraction of two counting measurements, scaled by systematic parameters. Typically, the Gaussian likelihood applies in most background-limited measurements or those measurements dominated by systematic uncertainties. The Poisson likelihood applies to background-free or very low statistics measurements. 

\begin{figure}
     \centering
     \includegraphics[width=\linewidth,trim={0 0 1cm 1cm},clip]{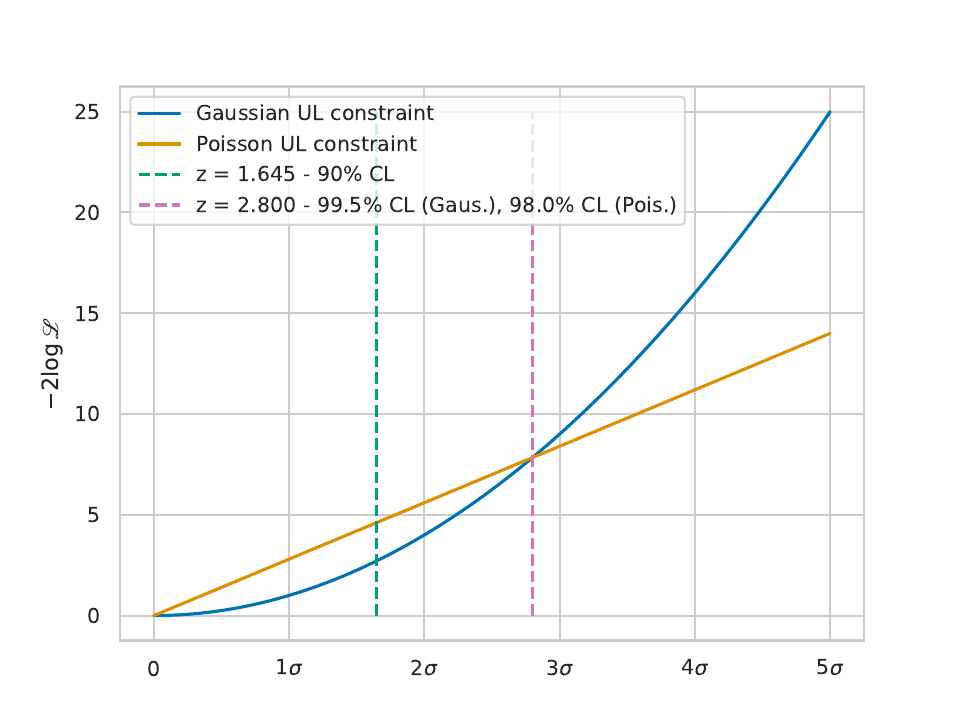}
    \caption{Comparison of Gaussian and Poisson likelihoods for an upper limit at 90\% CL with corresponding unit $\sigma$.}
    \label{fig:comparelikelihoods}
\end{figure}

In Figure \ref{fig:comparelikelihoods}, a limit is placed at 90\% CL for Gaussian and Poisson likelihoods with unit $\sigma$. The negative log-likelihood is shown with its minimum adjusted to zero in each case so as to reflect only the curvature. Note that, from the negative log-likelihood, the Gaussian constraint is looser below the 90\% CL, while Poisson constraint allows for a larger excursion from the limit. The $z$-score at which the two distributions have their negative log-likelihoods cross can be shown to be given by

\begin{equation}
    z = - \frac{\ln(1 - \textrm{CL})}{[\erf^{-1}(\textrm{CL})]^2} \cdot \textrm{UL}.
\end{equation}

Since in this work we are primarily concerned with using assay results to inform spectral fits or, similarly, projecting expected background rates, the Gaussian likelihood demonstrates the desirable behavior that very low activities are not strongly preferred, allowing for the possibility that higher activities near the reported upper limit are explored by the fit or projection. This additionally has the advantage of simplicity in that activities of materials and components are parameterized by only their mean and 1$\sigma$ uncertainty, even for those activities which are reported as an upper limit.

As explored in section \ref{sec:parttopart}, an average of several assays is often performed. In the case that multiple upper limits are reported for assays of the same part, we include in the average only the most stringent upper limit, further following the recommendation of Ref. \cite{mjdstatistics}.

\section{Incorporating part-to-part variation}\label{sec:parttopart}

Materials that make up a low-background experiment typically go through a rigorous process of selection, fabrication, and cleaning to achieve extremely radiopure parts. Still, the individual parts are subject to variation in the amount of contamination of introduced to them throughout these processes, or possibly introduced during handling or sample preparation for the assay itself. Some experiments have observed part-to-part variation of identical samples clearly exceeding the reported radioassay 1$\sigma$ uncertainty, (e.g. machined electroformed copper samples of \cite{Christofferson:2017nih, Abgrall:2016cct}). Typically, it is only possible to assay a subset of parts, or possibly only witness samples and not the parts themselves; this part-to-part variation therefore presents a challenge for the prediction of the activity of the in-use parts. Additionally, it is often not possible to discern whether this variation in assay results is due to variation in fabrication and cleaning of the part, and thus representative of the production chain, or if it is the result of different sample preparation and handling at the assay stage.

To account for variation in assay results of identical parts, one recent approach \cite{mjdstatistics} employs an uncertainty-weighted average of assay results with overall uncertainty inflated by the Birge ratio \cite{Birge_1932}, following the procedure of the Particle Data Group for combining measurements of the same parameter \cite{Workman:2022ynf}. Under the assumption that the parameter of interest has a single true value, in this case the activity of a part, this technique attempts to account for unconsidered and unreported systematic variance in different measurements of that parameter. Explicitly, given assay results results with central values $x_i$ and uncertainties $\delta_i$, the weighted average and its error are computed as

\begin{equation}
    \bar{x} \pm \delta_{\bar{x}} = \frac{\sum_i w_i x_i}{\sum_i w_i} \pm S \left(\sum_i w_i\right)^{-1/2},
    \label{eq:BirgeMethod}
\end{equation}

\noindent where $w_i = 1/(\delta_i)^2$ and S is the Birge ratio, given by
\begin{equation}
    S = \max \left(1, [\chi^2 / (N-1)]^{1/2} \right)
\end{equation}
(here $\chi^2 = \sum w_i (\bar{x} - x_i)^2$). That is, the Birge factor inflates all uncertainties by the reduced $\chi^2$.
Originally, this approach to a weighted average of measurements was intended to account for unknown systematic uncertainties that cause individual measurements of a quantity to have poor statistical agreement. By inflating the uncertainties of all measurements by the same scale factor, the approach is intended to be agnostic as to which measurements may be more or less accurate. In the context of radioassays, this approach is intended to reconcile differing specific activities as reported by different techniques or samples into an average activity and uncertainty that is reflective of the actual knowledge of the parts activity.

However, despite the suggestion that this approach can incorporate sample-to-sample variation, mathematically, this approach makes the fundamental assumption that a collection of parts or material should have a single specific activity; it does not formally account for the part-to-part variation that almost assuredly exists. Depending upon the relative scales of the part-to-part variation to the average part activity, this can have a large impact on the total extrapolated activity. To illustrate this, consider a collection of parts with individual activities drawn from a Gaussian distribution. If, unbeknownst to the experimentalist, an assay were conducted on a single part whose activity came from the near the extrema of the Gaussian distribution, a quite wrong conclusion about the parts' total activity would be drawn. Likewise, if several different assays were conducted on different parts from this same distribution, the experimentalist may employ the weighted average technique to estimate the average and inflate the uncertainty accordingly, which would approximate the underlying Gaussian distribution. However, this will systematically lead to an underestimate on the total activity and its variance, as will be shown later.

Another method that has been used to assign activities is an approach we label the ``independent batches" approach. In this method, an assay is conducted of each ``batch" of parts. Then the resulting activity is propagated directly to all parts from that batch. This ignores any part-to-part variance and assumes that any disagreement between assays is instead representative of the entire batch of parts. The performance of this method is dependent on the underlying distribution of activities, which may or may not be consistent with a Gaussian part-to-part variation.

Instead, we propose an improved method to more properly account for part-to-part variation, as well as differing assay results, which provides more accurate results in a wider range of part-to-part variances.

\subsection{Improved averaging method}

Let us consider that the individual parts in an experiment may have some variation in their activities, which can be due to differences in surface contamination, contamination introduced during machining or fabrication, or contamination from other sources. We consider here a single component made of several identical parts. As our model, we assume that each discrete part has an activity drawn from a Gaussian distribution with mean $\mu$ and variance $\upsilon^2$. We perform a number of assays $N_A$ of these parts, where each assay reports an activity with some resolution $x_i \pm \delta_i$, restricting the $x_i$ to be non-negative and taking the $\delta_i$ to be known. Then, in our experiment, we use a number of parts $N$, where usually $N_A \ll N$. Since it is usually computationally intractable to simulate and/or fit the activities of each individual part separately, we are interested in only the total activity of the parts. 

The task is therefore to predict the expected total activity using the assay results of a limited number of samples from our best estimates of the average activity, $\hat{\mu}$, and the part-to-part variation, $\hat{\upsilon}$. Under these assumptions, the 1$\sigma$ interval of the total activity of $N$ parts is given by

\begin{equation}\label{eq:totalactivity}
\eta = N \mu \pm \sqrt{N} \upsilon,
\end{equation}

\noindent where the variance comes from the random selection from the part to part activity distribution. Since the uncertainty on our estimate of $\hat{\mu}$, $\delta_{\hat{\mu}}$, will contain a $\upsilon$ term, this variance enters twice into the uncertainty of the total activity, though the importance of this will be explored. Note that we will ignore the uncertainty on $\upsilon$ as a small correction at sub-leading order.

Let us consider the result of an assay $x_i \pm \delta_i$. The variance of $x_i$ about $\mu$ is given by the sum of the measurement variances $\delta_i^2$ and the part-to-part variance $\upsilon^2$. The log-likelihood is thus

\begin{multline}
           -2 \ln \mathcal{L}(\{x_i\} | \mu, \upsilon, \{\delta_i\}) = \\
        \sum_{i=1}^{N_A} \left(\ln[2 \pi (\upsilon^2 + \delta_i^2)] + \frac{(x_i - \mu)^2}{\upsilon^2 + \delta_i^2}\right). 
\end{multline}

\noindent The maximum likelihood estimates of $\upsilon$ and $\mu$ are found by solving

\begin{multline}
    \pdv{\upsilon} (-2 \ln \mathcal{L}(\{x_i\} | \mu, \upsilon, \{\delta_i\}))  =\\ 
    \sum_{i=1}^{N_A} \left(\frac{2 \upsilon}{\upsilon^2 + \delta_i^2} - 2 \upsilon \frac{(x_i - \mu)^2}{(\upsilon^2 + \delta_i^2)^2}\right) = 0.
\end{multline}

\noindent and

\begin{equation}
     \pdv{\mu} (-2 \ln \mathcal{L}(\{x_i\} | \mu, \upsilon, \{\delta_i\})) = \sum_{i=1}^{N_A} \frac{2 (\mu - x_i)}{\upsilon^2 + \delta_i^2} = 0.
\end{equation}

\noindent For $\hat{\mu}$, the solution is 

\begin{equation}\label{eq:hatmu}
    \hat{\mu} = \left(\sum_{i=1}^{N_A}\frac{x_i}{\upsilon^2 + \delta_i^2}\right) \cdot \left(\sum_{i=1}^{N_A}\frac{1}{\upsilon^2 + \delta_i^2}\right)^{-1},
\end{equation}

\noindent which is just a weighted average. The uncertainty on $\hat{\mu}$ is estimated by 

\begin{equation}
    \delta_{\hat{\mu}} = \left(\sum_{i=1}^{N_A} \frac{1}{\upsilon^2 + \delta_i^2}\right)^{-\frac{1}{2}},
\end{equation}

\noindent the typical uncertainty estimate for a weighted average (an average of the weights). Note here that uncertainty on $\mu$ is a weighted average of the $\delta_i$, each inflated (in quadrature) by the same additive constant $\upsilon$, in contrast to the Birge ratio treatment where the terms are inflated by a multiplicative constant instead. Note also that as $\upsilon \rightarrow 0$, this becomes just the standard weighted average.

We note that $\hat{\upsilon}$ does not have an analytic solution; since a simple expression is desired, we instead make an approximation. To estimate $\upsilon^2$, we first approximate the variance of the $\{x_i\}$ as 

\begin{equation}
    \textrm{var}(\{x_i\}) = \frac{ \sum_{i=1}^{N_A} (x_i - \bar{x})^2}{N_A - 1} \approx \upsilon^2 + \frac{1}{N_A}\sum_{i=1}^{N_A} \delta_i^2,
\end{equation}

\noindent which becomes exact in the limit of large $N_A$. Here, $\bar{x}$ is the direct mean calculated from the assays and the second term on the right-hand side is the average assay resolution, $\bar{\delta_i}^2$. We can then estimate $\upsilon^2$ as follows, with the restriction that it should be non-negative in the case that there is no evidence for a part-to-part variation within the resolution uncertainty,

\begin{equation}\label{eq:upsilon2}
   \upsilon^2 \approx \max\left(0, \frac{ \sum_{i=1}^{N_A} (x_i - \bar{x})^2}{N_A - 1} - \frac{1}{N_A}\sum_{i=1}^{N_A} \delta_i^2\right).
\end{equation}

Using the estimate of $\upsilon^2$ from Equation \ref{eq:upsilon2}, we can estimate $\hat{\mu}$ from Equation \ref{eq:hatmu}. We can then use our calculated values for $\hat{\mu}$, $\delta_{\hat{\mu}}$, and $\upsilon$ to compute the expected total activity from Equation \ref{eq:totalactivity}, where we also include the uncertainty on our estimate of the mean.
In this approximation, the uncertainty on the total activity $\eta$ is

\begin{equation}
\eta \approx N (\hat{\mu} \pm \delta_{\hat{\mu}}) \pm \sqrt{N} \upsilon
\label{eq:AVmethod}
\end{equation}

\noindent where we ignore the uncertainty on $\upsilon$ as a small correction. We refer to this estimate of the total activity as the ``additive-variance'' estimate.

To explore the dependence on $N$ and $N_A$, we note that when the $\delta_i$ are similar such that $\delta_{\hat{\mu}} \approx (\upsilon^2 + \bar{\delta}_i^2 ) / N_A$, we find approximately

\begin{equation} \label{eq:expand}
  \eta   \approx N \hat{\mu} \pm \frac{N}{\sqrt{N_A}} \bar{\delta_i} \pm \frac{N}{\sqrt{N_A}} \upsilon \pm \sqrt{N} \upsilon.
\end{equation}

Notably, the last term that provides the ``extra'' variance in Eq.~\ref{eq:totalactivity} will only dominate when the number of assays is large compared to the number of parts. As this is rarely, if ever, the case for low-background experiments, the uncertainty will usually be dominated by that of the assays. 

For many assays, results are reported as a specific activity (activity per unit mass). Additionally, in most ICP-MS assays, only a small amount of material from a component or part is assayed. These results need some care in interpretation under this approximation, as this has been derived assuming an integer number of discrete parts. Assuming for simplicity that each assay samples the same mass $m$, we can consider that the activity of the sample is drawn from a distribution with mean $m \mu$ and variance $(m \upsilon)^2$, where $\mu$ and $\upsilon$ are now specific activities. This is equivalent to dividing a full component with mass $M$, typically on the order of ~1 kg, into $N$ identical parts of mass $m$, typically $<$1 g, a small subset of which is assayed. This implies that the ``extra" variance contributes negligibly due to the extremely large number of parts in this interpretation, as seen from expanding Eq. \ref{eq:AVmethod},

\begin{equation} \label{eq:expandmass}
  \eta   \approx M (\hat{\mu} \pm \delta_{\hat{\mu}} ) \pm \sqrt{M} \sqrt{m} \upsilon.
\end{equation}

\noindent It is straightforward to extend this method to parts with differing masses.

The key difference between the aforementioned Birge-factor approach and the proposed additive variance approach is the treatment of the common part-to-part variance. In the Birge-factor approach, this is included through a multiplicative scale factor. In the new treatment, however, this is an {\it additive} contribution (in quadrature). In the case where we have several assays with resolutions that vary by an order of magnitude or more, if there is also large variance on the sample activities, multiplying the uncertainties by a common scale factor may not sensibly account for the part to part variance. Adding a common part-to-part variance instead can perform much more sensibly.

\begin{figure*}
    \centering
     \includegraphics[width=\textwidth,trim={1.5cm 0 2cm 1cm},clip]{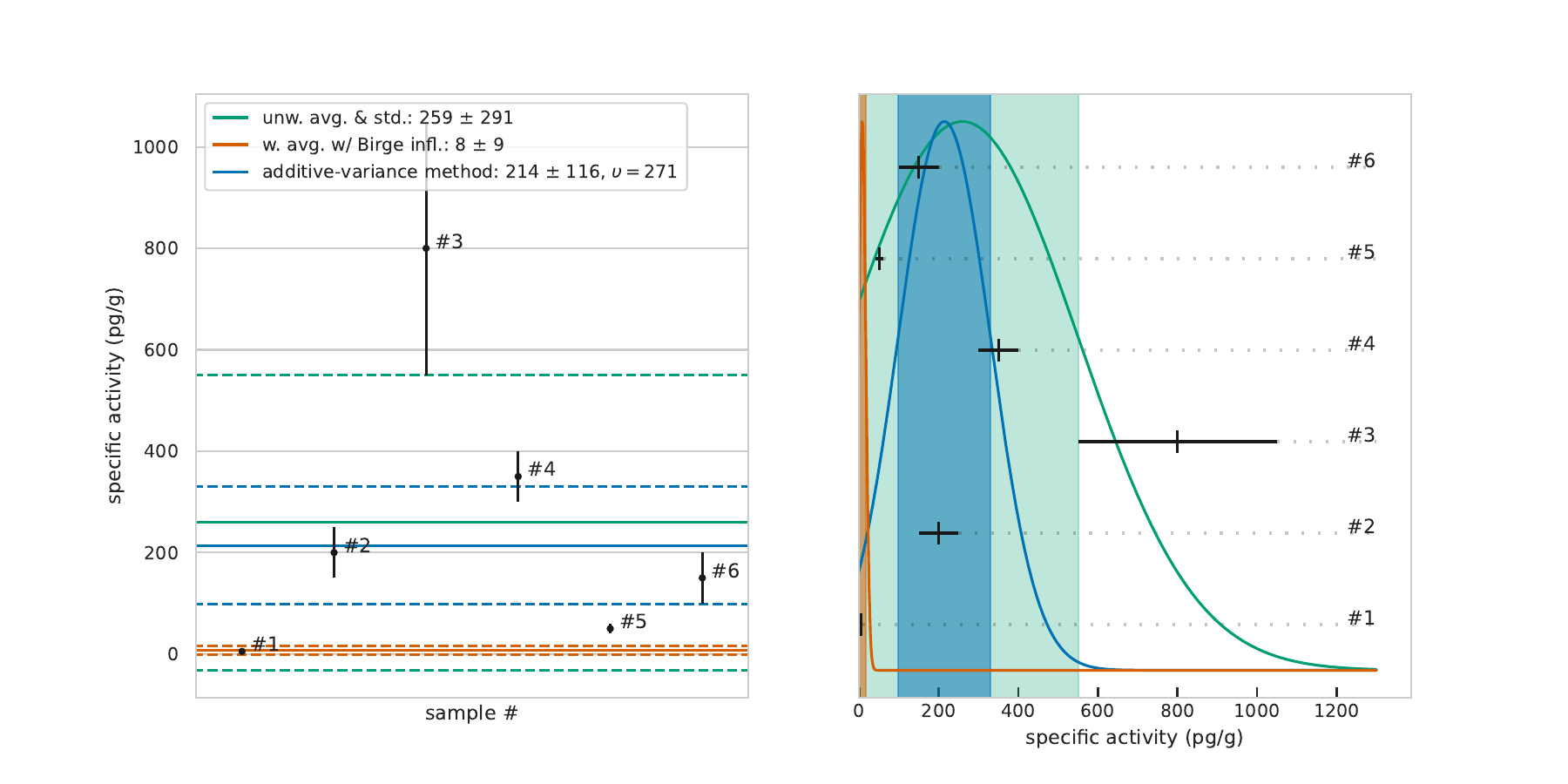}
    \caption{ Hypothetical results of ICP-MS assay of six samples of the same part, inspired by real-world results. The results and uncertainties of the six assays are shown, as well as the resulting mean and uncertainty of the additive-variance procedure (Eq.~\ref{eq:AVmethod}, blue) and the previous procedure of weighted average with Birge ratio inflation (Eq.~\ref{eq:BirgeMethod}, orange). We compare also to an unweighted average and standard deviation (green), ignoring the individual uncertainties on the results. An ideogram \cite{Workman:2022ynf} is shown in the right panel to further highlight the difference.}
    \label{fig:illustration}
\end{figure*}

This difference is illustrated in Figure \ref{fig:illustration} where we look at a hypothetical case of ICP-MS assays of identical parts inspired by real-world results. In our example, six assays were conducted in total.
These results have large variance and very large differences in assay resolution, an occurrence which is occasionally reported in assay reports available to us. One could consider removing outliers from the averaging procedure, but since handling was performed identically among these samples, it is unclear which results should be considered outliers.

\begin{table}
\renewcommand{\arraystretch}{1.2}
\caption{Hypothetical ICP-MS assay results for six samples of an identical part.}
\begin{tabularx}{\linewidth} {>{\centering\arraybackslash}X>{\centering\arraybackslash}X}
\hline
sample \# & ICP-MS result (pg/g)\\
\hline
1 & 5 $\pm$ 2\\
2 & 200 $\pm$ 50\\
3 & 800 $\pm$ 250\\
4 & 350 $\pm$ 50\\
5 & 50 $\pm$ 10\\
6 & 150 $\pm$ 50\\
\hline
\hline
direct average and standard deviation & 259 $\pm$ 291\\
weighted average with Birge ratio inflation & 8 $\pm$ 9\\
additive-variance method & $\mu = 214 \pm 116$, $\upsilon = 271$\\
\hline
\end{tabularx}
\label{tab:beamenergies}
\end{table}

Applying a weighted average with Birge ratio inflation predominantly assigns weight to the assay with the higher resolution, which coincidentally (or not) has the smallest activity. Applying either an unweighted, direct average (and unweighted sample standard deviation) or the proposed additive-variance method returns an average activity an order of magnitude larger in this example, with larger uncertainties corresponding better with what one would intuitively expect. Compared to the unweighted average and sample standard deviation, the uncertainty from the additive-variance method is smaller and  appropriately accounts for the resolutions of the different measurements.

\subsection{Comparison of approaches}

We run a series of Monte Carlo (MC) simulations to compare these three approaches to averaging; the weighted average with Birge ratio inflation, the independent batches approach, and the proposed additive-variance average using Eq. \ref{eq:AVmethod}. For each MC, we specify the true unknown values $(\mu_t, \upsilon_t)$, as well as the number of assays $N_A$ and number of parts $N$. We draw an activity for each part and assay from a Gaussian distribution with mean $\mu_t$ and variance $\upsilon_t^2$, truncated at zero such that no negative activities are allowed; truncation of the distribution becomes important as $\upsilon / \mu \rightarrow 1$. 

Assays are performed destructively in the MC; this is analogous to an ICP-MS assay or gamma assay of witness parts not used in the final experiment. In addition, we select a resolution $\sigma$ for each assay; the resolutions are drawn from a folded Gaussian with a provided mean and variance. The exact distribution of resolutions should not matter, but these two parameters allow for an easy way to specify how disparate assay resolutions should be in a given trial.

Once we have drawn the activity for each assayed part and its resolution, we draw a result of the assay from a Gaussian centered at the true value of the activity with variance given by the resolution, truncated at zero such that no negative values are drawn. We take the resolution of the assay as the uncertainty on the assay measurement.  The specific distribution from which we draw the assay result may not reflect the assays performed in reality; this is only an approximation. 

\begin{figure*}
     \centering
    
         \includegraphics[width=\linewidth,trim={1.5cm 0 2cm 1cm},clip]{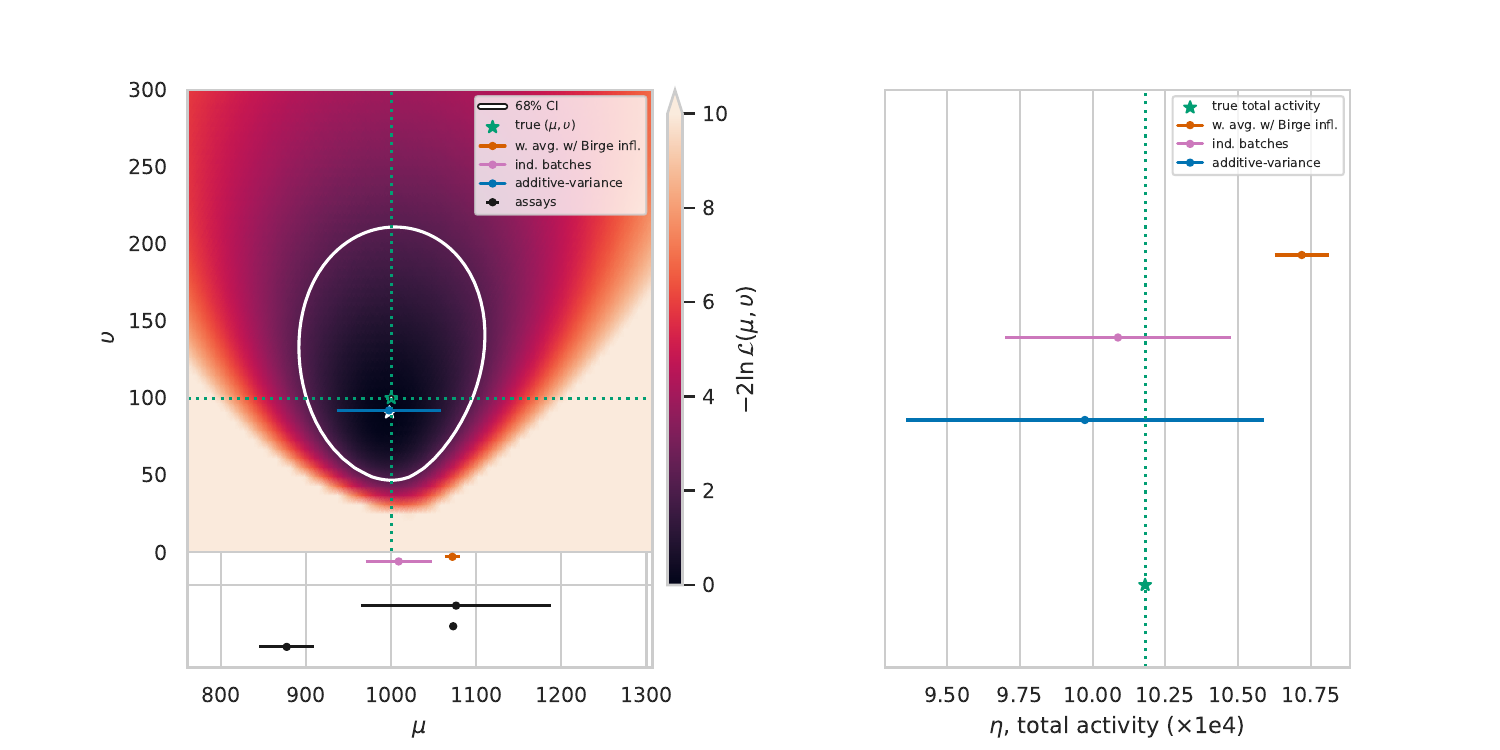}
         
    \caption{Example of a pseudo-experiment with parameters $N_A=3, N_C=100, \nu/\mu = 0.1,$ and $ \sigma/\mu = 0.1 \pm 0.1$. The left figure shows the $-2 \ln \mathcal{L}(\mu, \upsilon)$ surface from the assays while the right figure shows the total activity of the parts. The assay results and their uncertainties are shown at the bottom of the left figure, while the colored points with error bars show the results of the different statistical interpretations: weighted average with Birge ratio inflation in orange, independent batch treatment in pink, and our new method in blue. The true value is shown in green. In this particular example, one of the assays has much lower uncertainty than the others.}
    \label{fig:trialexamples}
\end{figure*}

In this way, we fully simulate the part-to-part variance as well the assay measurement resolution. Using these assay results, we apply the above approaches and estimate the total activity for the $N$ parts. In Figure \ref{fig:trialexamples}, we show some examples from the MC trials to demonstrate how the various methods perform in particular pseudo-experiments. The plots on the left show the $-2 \ln \mathcal{L}(\mu, \upsilon)$ surface with 68\% confidence level (CL) and most likely value shown. The assay results for this trial are shown in the lower portion of the plot. The results from each of the three methods are also shown; we consider that each assay corresponds to a separate batch for the independent batch approach such that an equal fraction of the parts are attributed to each assay. On the right side, the plots show the predicted total activity with 68\% CL indicated compared to the true total activity for the trial. Each of these examples uses 3 assays and 100 parts, which are similar to modern real-world experiments. The part-to-part variation $\upsilon_t$ was taken as 10\% of the average activity $\mu_t$, i.e. $\upsilon / \mu = 0.1$. The assay resolution was also taken as 10\% of $\mu_t$, with standard deviation of the assay resolution of 10\% of $\mu_t$, which we notate as $\sigma / 
\mu = 0.1 \pm 0.1$. This assay resolution distribution therefore allows for large excursions from the average assay resolution, including to very low uncertainties, as demonstrated in the example in Figure \ref{fig:trialexamples}. 

To quantitatively compare the three averaging methods, we choose the frequentist coverage as a metric, which characterizes the accuracy of the reported error bars. Although coverage is guaranteed to match or exceed the chosen confidence level for a full Neyman construction, each of the three averaging methods discussed here employ approximations for simplicity and so are not expected to respect coverage in all cases. Evaluation of coverage can therefore elucidate the parameter regimes in which each method is expected to perform better or worse.

\begin{figure*}[h! tb]
     \centering
    
    \includegraphics[width=\linewidth,trim={1.5cm 0 2cm 1cm},clip]{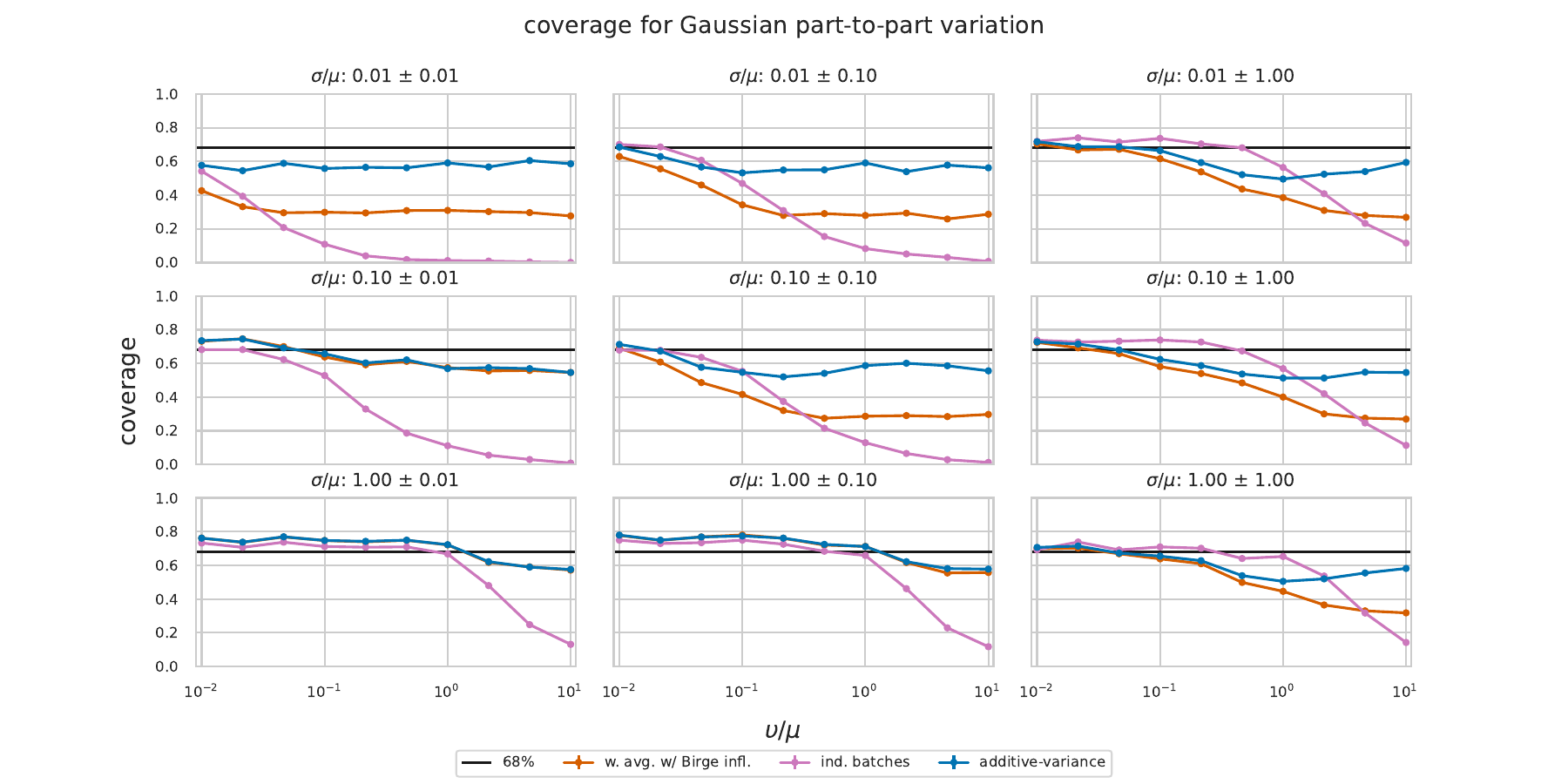}

    \caption{Coverage results from pseudo-experiments with the three methods with $N = 100$ and $N_A = 3$ with a Gaussian distributed part-to-part variation for the parts: weighted average with Birge ratio inflation in orange, independent batch approach in pink, and the additive-variance method in blue.}
    \label{fig:coverage3ppv}
\end{figure*}

\begin{figure*}[h! tb]
     \centering
    
    \includegraphics[width=\linewidth,trim={1.5cm 0 2cm 1cm},clip]{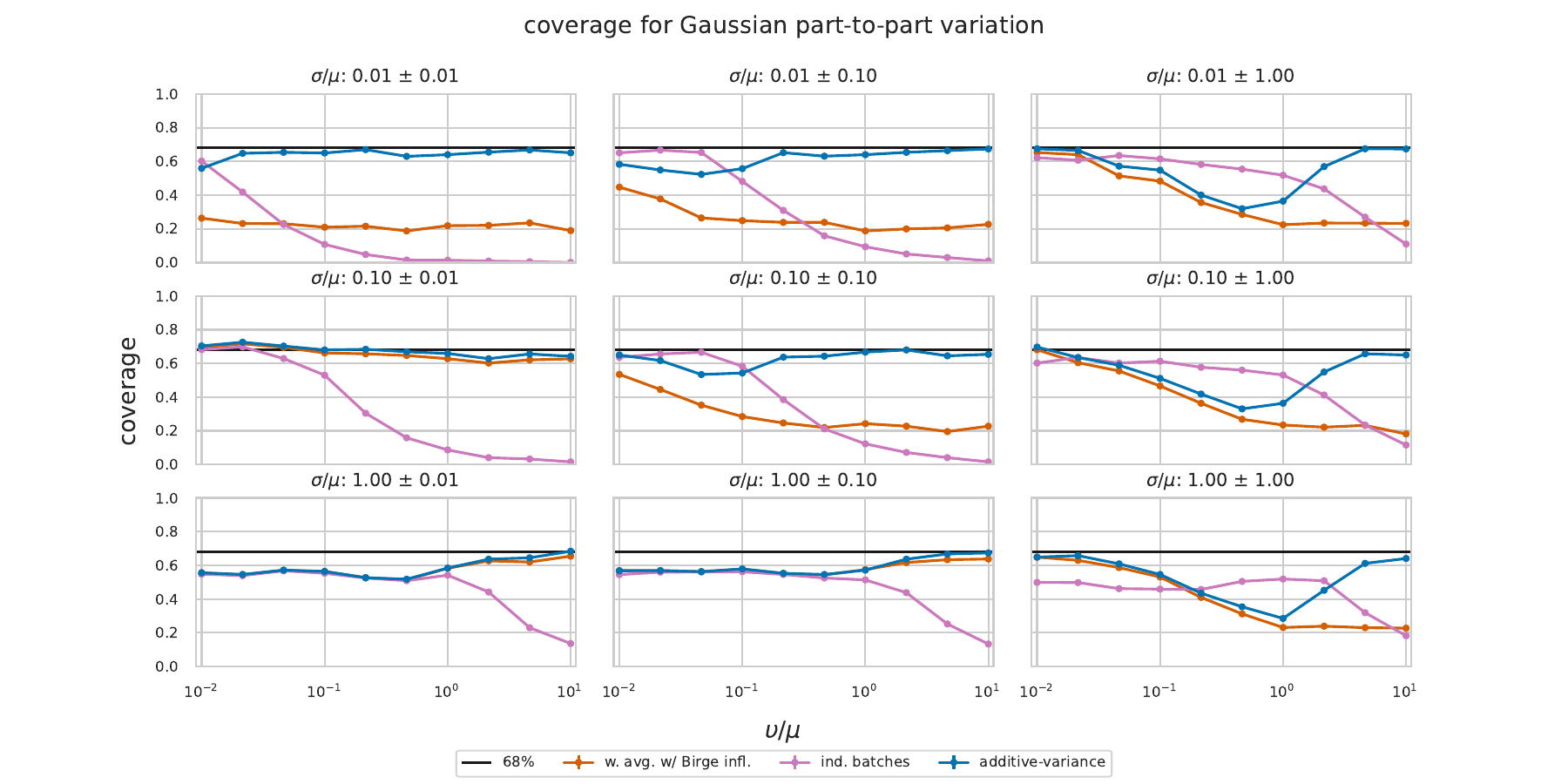}

    \caption{Coverage results from pseudo-experiments with the three methods with $N = 100$ and $N_A = 10$ with a Gaussian distributed part-to-part variation for the parts: weighted average with Birge ratio inflation in orange, independent batch approach in pink, and the additive-variance method in blue}
    \label{fig:coverage10ppv}
\end{figure*}

We compare the coverage of the three averaging methods under different $\mu_t, \upsilon_t$, and $\sigma$. We set $N=100$ and show results for $N_A = 3, 10$ in Figures \ref{fig:coverage3ppv} and \ref{fig:coverage10ppv}, respectively. Since we work under a Gaussian approximation, we parameterize the problem in terms of the ratios $\upsilon/\mu$ and $\sigma/\mu$, which allows us to make coverage statements about the whole parameter space independent of the particular $\mu$, which we choose for convenience to be $\mu_t=1000$. This is possible since poor coverage is a result of the boundary of the parameter space at $\mu=0$ that restricts activities to be physically possible. The fraction of the Gaussian probability density that is excluded by the physical boundary is controlled only by the relative size of $\upsilon/\mu$ and $\sigma/\mu$. Therefore, the horizontal axes have units of $\upsilon/\mu$ to show the coverage as the part-to-part variation becomes large compared to its average. The labels above each plot provide the resolution mean and resolution variation relative to $\mu_t$. Each point is the result of 1000 pseudo-experiments; statistical uncertainties on the coverage are on the order of $\sim$ 1\%.

The motivating case for this work is when the part-to-part variance is larger than the assay resolution; this corresponds to plots in the upper left of the grid. In this case, the coverage of the additive-variance method proposed here is much better than previous methods. The weighted average with Birge ratio inflation significantly undercovers in many scenarios, as does the independent batches approach. 

The undercoverage of the additive-variance method at various points may be due to several factors. One issue is that the underlying distributions are close to the boundary of the parameter space and are not truly Gaussian, while the averaging method assumes an underlying Gaussian distribution. At low $N_A$, bias in the estimation of the part-to-part standard deviation may still be large, resulting in undercoverage. This may explain why the coverage does not quite rise to the 68\% CL in the top left corner plot for $N_A = 3$, where the assay resolution is very small. Surprisingly, there is also significant undercoverage in the case of $N_A = 10$ for large assay resolutions. This is likely due to the implemented sampling scheme which draws activities from a truncated Gaussian distribution, artificially reducing the variance of the assay results, more than would be expected from the part-to-part variance. At high part-to-part variance, the underlying probability density function (PDF) for the true part values is a truncated Gaussian, so that agreement with the actual variance is restored.

The additive-variance method has much improved performance compared to the previous methods, under the assumption of a Gaussian distributed part-to-part variance. For further comparison under different scenarios, we also consider a few alternative treatments at various stages of the MC: 
\begin{itemize}
    \item For the underlying distribution of activities, we also consider a true batch-to-batch variation such that all parts from a particular batch have the same activity. We take the number of batches to be the number of assays, where one assay of each batch is conducted. The activities of the batches are then drawn from a truncated Gaussian distribution with mean $\mu_t$ and variance $\upsilon_t^2$.

    \item At the assay level, we alternatively implement a level of detection treatment where if an assay results in an activity $<3\sigma$, a 99.7\% C.L. upper limit of $3\sigma$ is instead reported. The limit is then incorporated into the different averages per the treatment described in section \ref{sec:basics}, i.e. converted to a measurement of $0 \pm \sigma$.

    \item Accordingly, in the case that more than 1 upper limit is reported, we consider the case of using all upper limits and of using only the most stringent upper limit in the different averages.
\end{itemize} 

\begin{figure*}[h! tb]
     \centering
    
    \includegraphics[width=\linewidth,trim={1.5cm 0 2cm 1cm},clip]{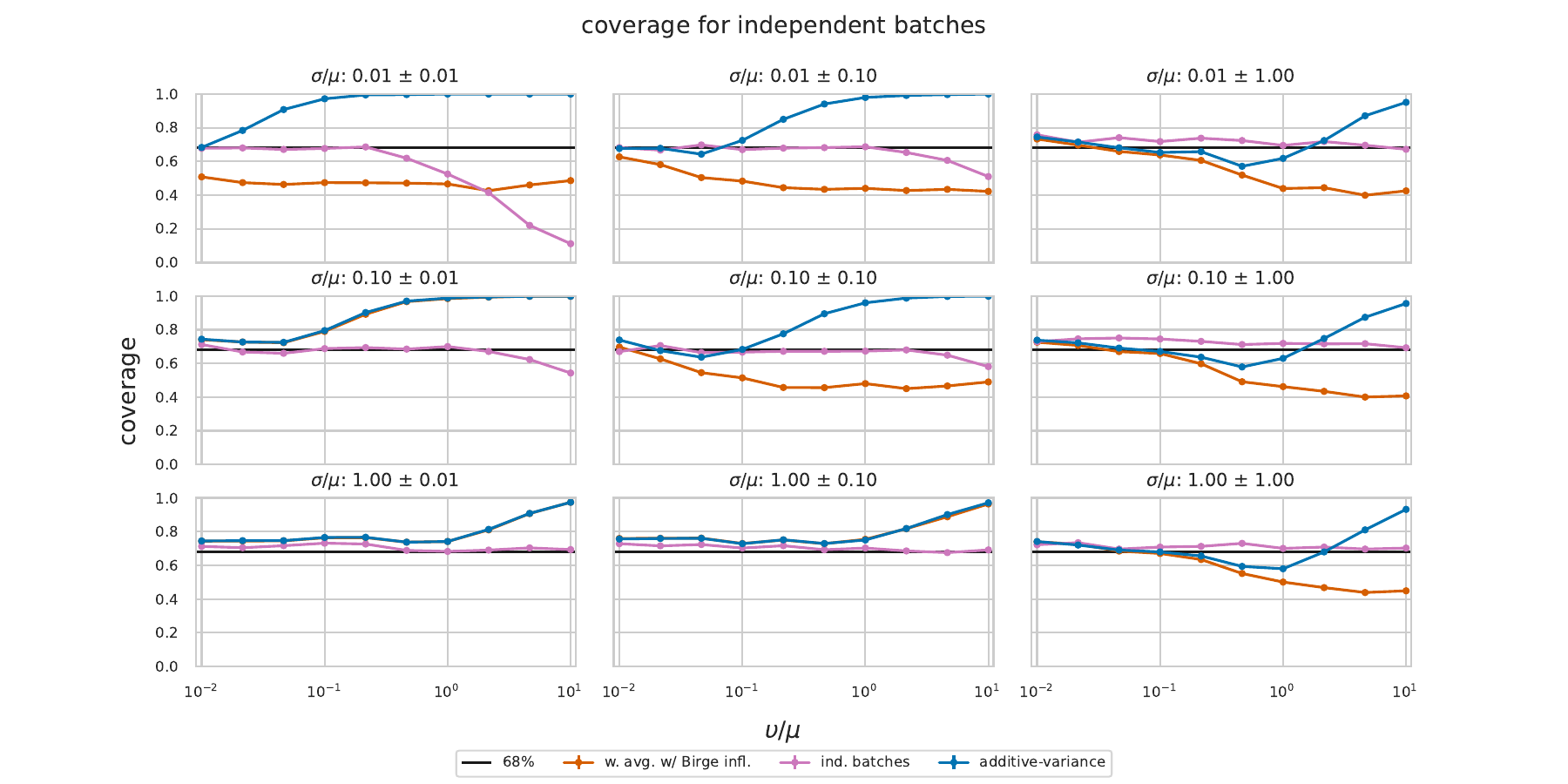}

    \caption{Coverage results from pseudo-experiments with the three methods with $N = 100$ and $N_A = 3$ with part activities from three independent batches: weighted average with Birge ratio inflation in orange, independent batch approach in pink, and the additive-variance method in blue.}
    \label{fig:coverage3indbatch}
\end{figure*}

\begin{figure*}[h! tb]
     \centering
    
    \includegraphics[width=\linewidth,trim={1.5cm 0 2cm 1cm},clip]{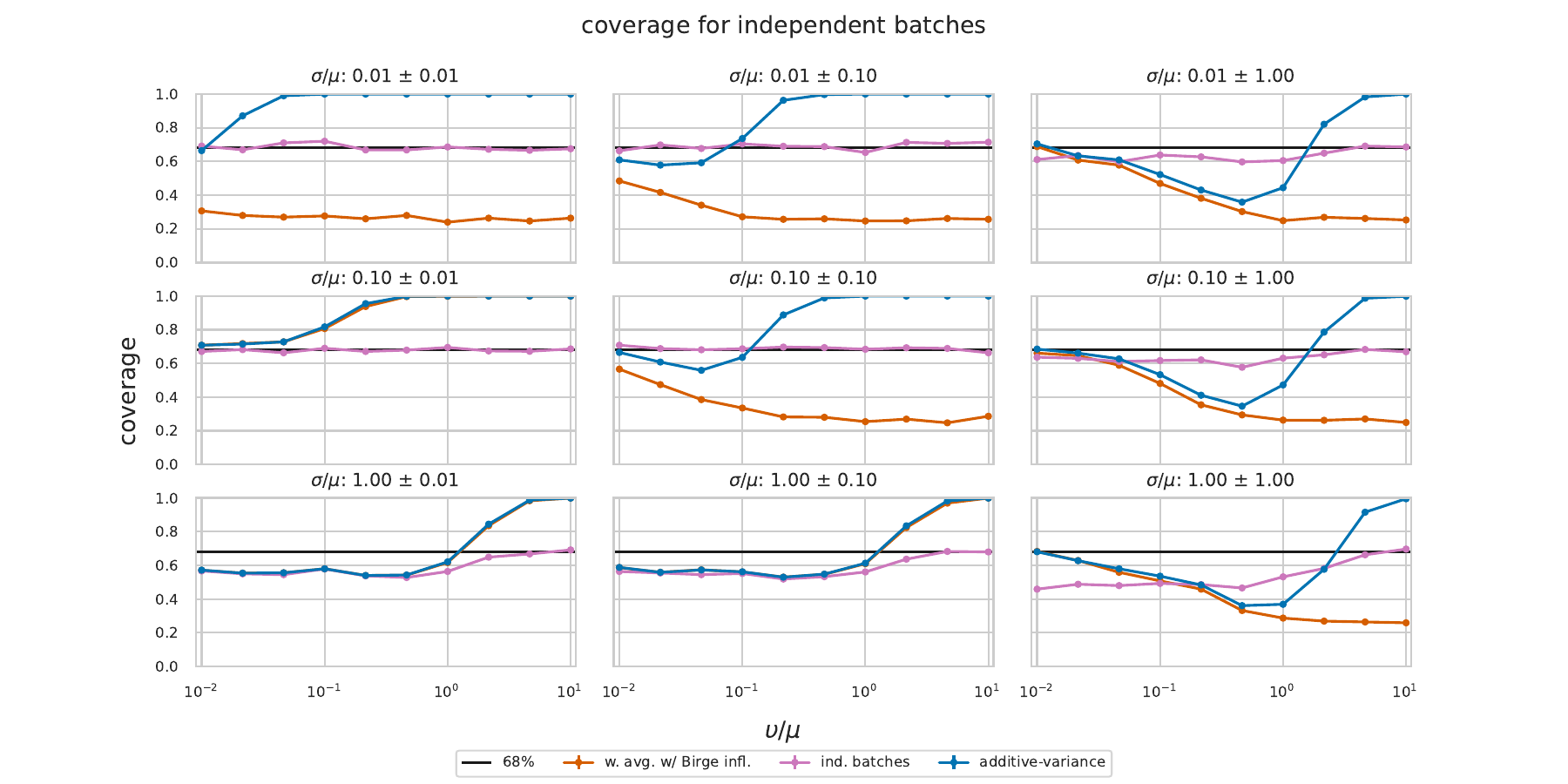}

    \caption{Coverage results from pseudo-experiments with the three methods with $N = 100$ and $N_A = 10$ with part activities from ten independent batches: weighted average with Birge ratio inflation in orange, independent batch approach in pink, and the additive-variance method in blue}
    \label{fig:coverage10indbatch}
\end{figure*}

Figures \ref{fig:coverage3indbatch} and \ref{fig:coverage10indbatch} show the coverage with the true part distribution coming from three and ten, respectively, independent batches and corresponding assays. Here, the independent batches approach generally has correct coverage, as would be expected. The weighted average with Birge ratio inflation significantly undercovers in this case. The additive-variance method significantly overcovers in most cases since it ascribes higher variance than in reality. However, since we in general cannot {\it a priori} know which distribution parts may be drawn from, the overcoverage of the additive-variance method in this instance seems an acceptable compromise.

\begin{figure*}[h! tb]
     \centering
    
    \includegraphics[width=\linewidth,trim={1.5cm 0 2cm 1cm},clip]{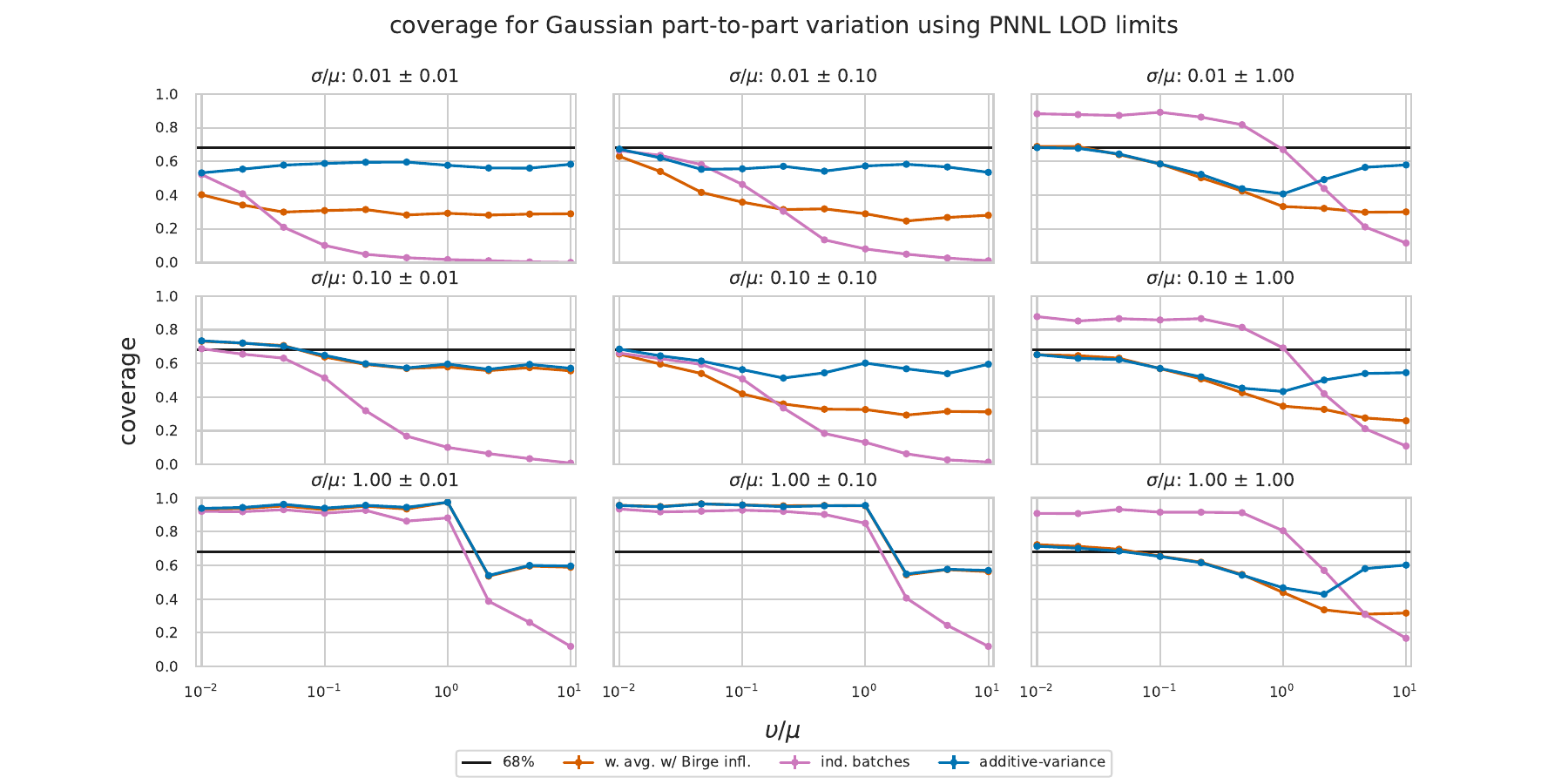}

    \caption{Coverage results from pseudo-experiments with the three methods with $N = 100$ and $N_A = 3$ with a Gaussian distributed part-to-part variation for the parts, using upper limits for assays not meeting a 3$\sigma$ level of detection. The weighted average with Birge ratio inflation is in orange, independent batch approach is in pink, and the additive-variance method is in blue.}
    \label{fig:coverage3lod}
\end{figure*}

\begin{figure*}[h! tb]
     \centering
    
    \includegraphics[width=\linewidth,trim={1.5cm 0 2cm 1cm},clip]{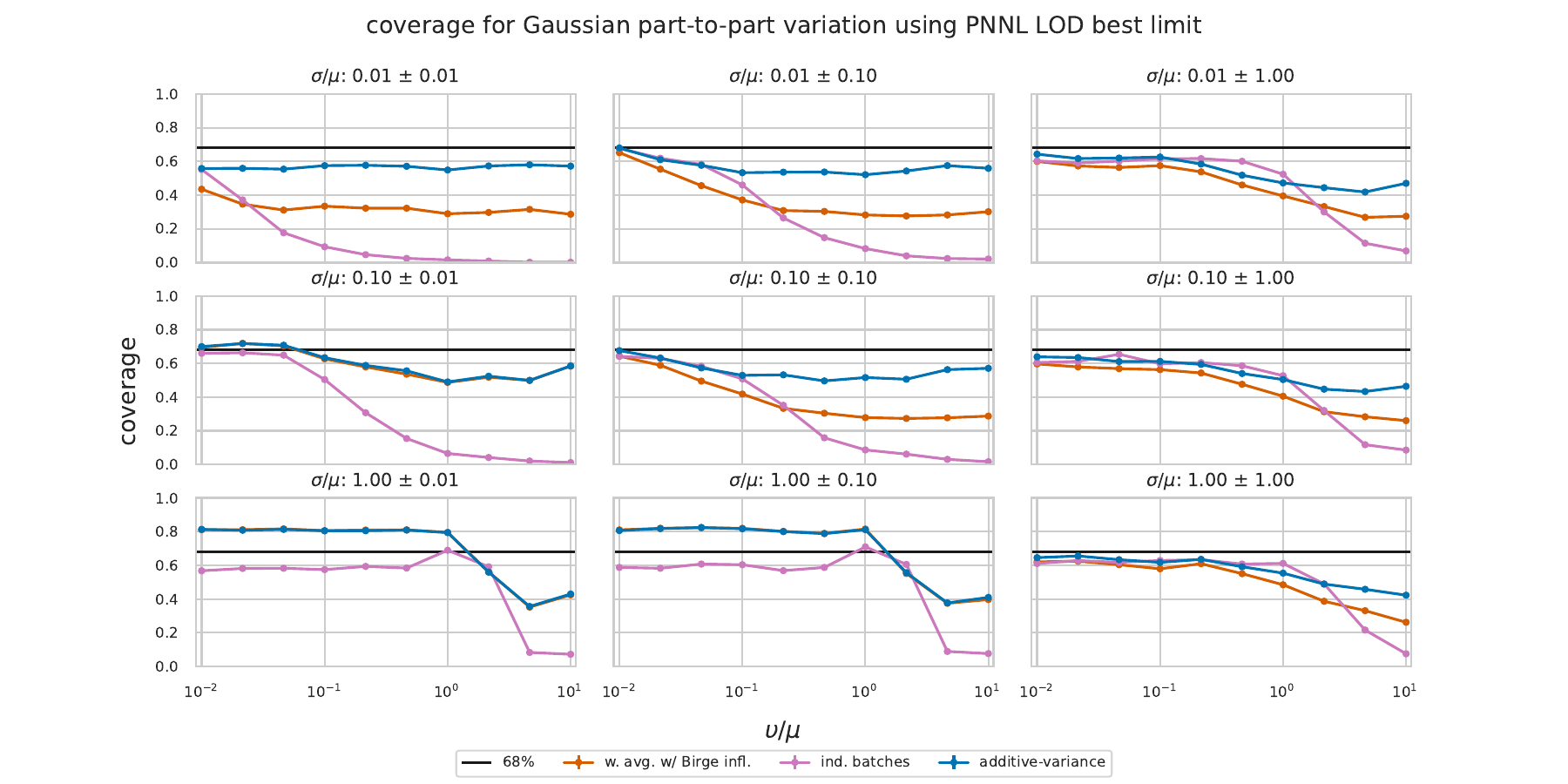}

    \caption{Coverage results from pseudo-experiments with the three methods with $N = 100$ and $N_A = 3$ with a Gaussian distributed part-to-part variation for the parts, using upper limits for assays not meeting a 3$\sigma$ level of detection. Only the most stringent limit is used, with other upper limits discarded. The weighted average with Birge ratio inflation is in orange, independent batch approach is in pink, and the additive-variance method is in blue.}
    \label{fig:coverage3lodbest}
\end{figure*}

Finally we consider the case where upper limits are reported for assays not meeting a level-of-detection threshold of $3\sigma$ as described previously, using the Gaussian distributed part-to-part variation for the underlying distribution. Further, we consider the cases of including all limits or only the most stringent. The results are shown in Figures \ref{fig:coverage3lod} for the inclusion of all upper limits and \ref{fig:coverage3lodbest} for the inclusion of only the most stringent upper limit. In general, using only the most stringent upper limit gives coverage closer to the expected value.

\section{Adding component activities}\label{sec:addcomponents}

Finally, we make a note on the addition of component activities into a single component group for e.g. background simulations. This a routine requirement for experiments to avoid the complexity of simulating every component or including all components separately in a fit to data. It is not always obvious how one should propagate the assay measurements and uncertainties of these different components into the total activity of the group. We consider a simplified case where these components contribute a nearly identical background PDF to the overall experiment, usually due to physical proximity. It is often therefore preferable to simulate them as a single component group to reduce computation time and analysis complexity. It is straightforward to extend this treatment to components with different background PDFs that include e.g. energy-dependence, but we consider only the simplified case here for clarity.

Let us consider two components $A$ and $B$ with true activities $\mu_A$ and $\mu_B$. These component activities sum to form a single component group activity $\mu_{\mathrm{tot}} = \mu_A + \mu_B$ for which we simulate decays uniformly and create a PDF for our fit to our data, $d_{\mathrm{data}}$, which is often an energy histogram. Note that our data are only sensitive to $\mu_{tot}$, not to the individual $\mu_A$ and $\mu_B$. Let us assume that the components have been assayed, and we have appropriately considered the part to part variation as discussed in section \ref{sec:parttopart} as well as converted from upper limits to measurements of zero with some uncertainty per section \ref{sec:basics}. 

\begin{figure}[htb]
     \centering
         \includegraphics[width=\linewidth]{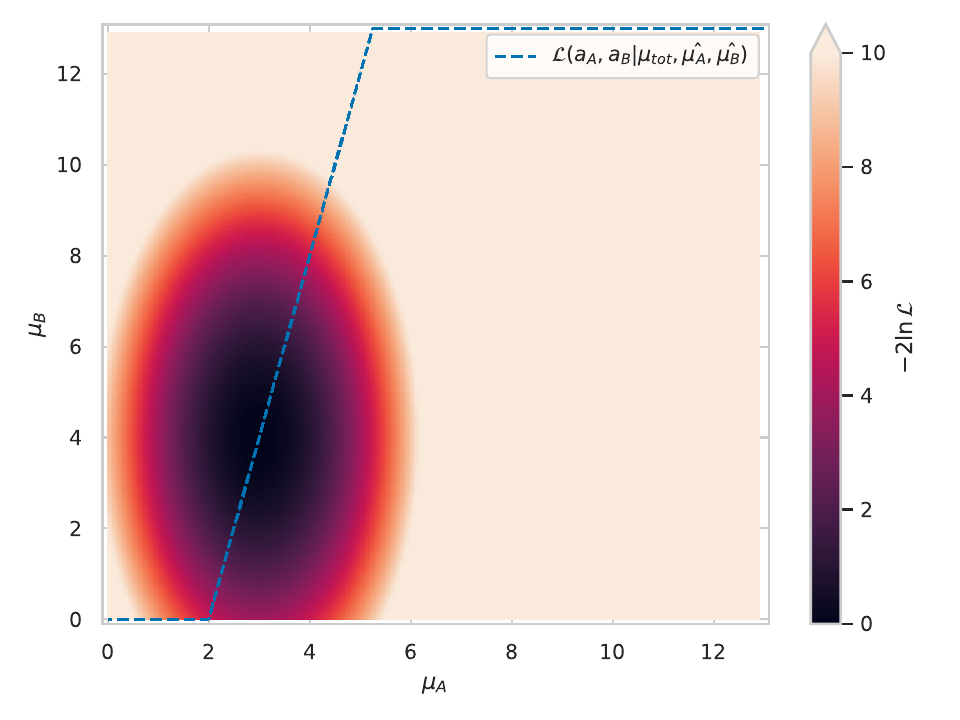}   
         \includegraphics[width=\linewidth]{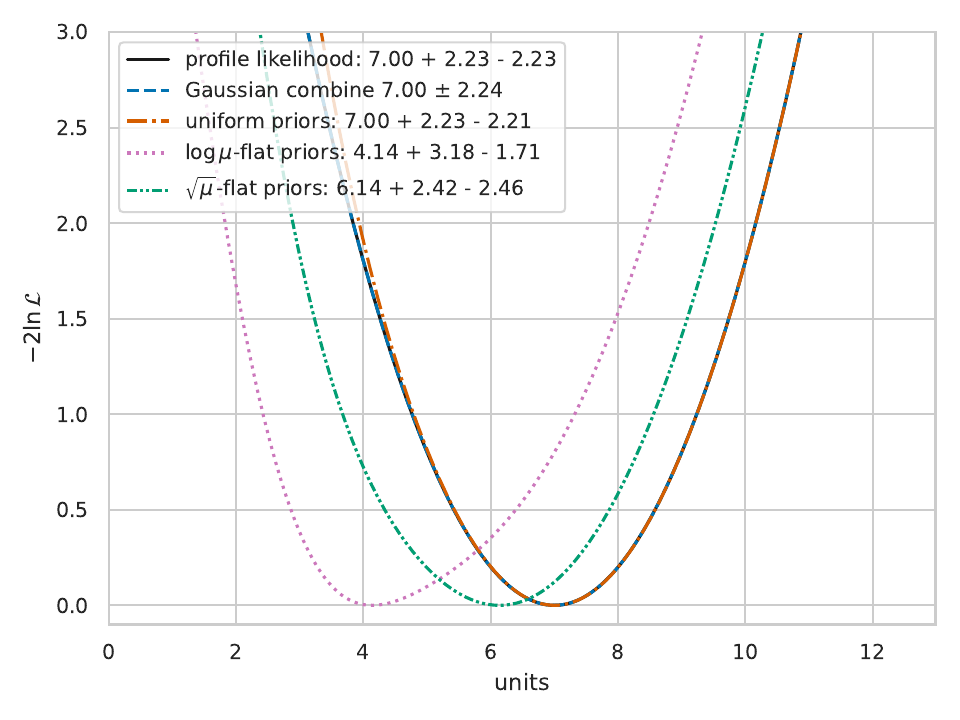}
    \caption{Comparison of profile likelihood and posterior constructions for an example with $a_A = 3 \pm 1$ and $a_B = 4 \pm 2$. In this case, where the results are far from the boundary at zero, the Gaussian approximation, profile likelihood, and posterior using uniform priors all agree. Other choices of prior result in different distributions.}
    \label{fig:profileexampleboring}
\end{figure}

The assay results $a_A$ and $a_B$ are typically reported with a 1$\sigma$ uncertainty which we take as the usual 68\% confidence interval (CI) and assume a Gaussian likelihood centered at the measured activity. Note that the true activities $\mu_A$ and $\mu_B$ cannot be negative. In the case where the confidence intervals are completely or almost completely positive, to find the sum of the activities of the two components, the usual method of adding the means and adding the 1$\sigma$ bands in quadrature clearly applies. This is shown in the example of Figure \ref{fig:profileexampleboring}, where we also demonstrate the profile likelihood method next described.

As an example, let us take the case where the assays have measured $a_A = 0 \pm 2$ (or $a_A < 3.3$ at 90\% CL) and $a_B = 4 \pm 2$ in some particular units. To combine the activities in this case, where one of the components has an activity strongly consistent with zero, we must be more careful in our construction of the total activity. From our data, the likelihood for the component's activity is given by 

\begin{equation}
\begin{split}
&\mathcal{L}(d_{\mathrm{data}}, a_A, a_B | \mu_\mathrm{tot}, \mu_A, \mu_B) \\&= \mathcal{L}(d_{\mathrm{data}} | \mu_\mathrm{tot}(\mu_A, \mu_B)) \cdot \mathcal{L}(a_A, a_B | \mu_{\mathrm{tot}}, \mu_A, \mu_B)\\
&= \mathcal{L}(d_{\mathrm{data}} | \mu_\mathrm{tot}(\mu_A, \mu_B)) \cdot \mathcal{L}(a_A | \mu_A) \cdot \mathcal{L}(a_B | \mu_B).
\end{split}
\end{equation}

\begin{figure}[htb]
     \centering
         \includegraphics[width=\linewidth]{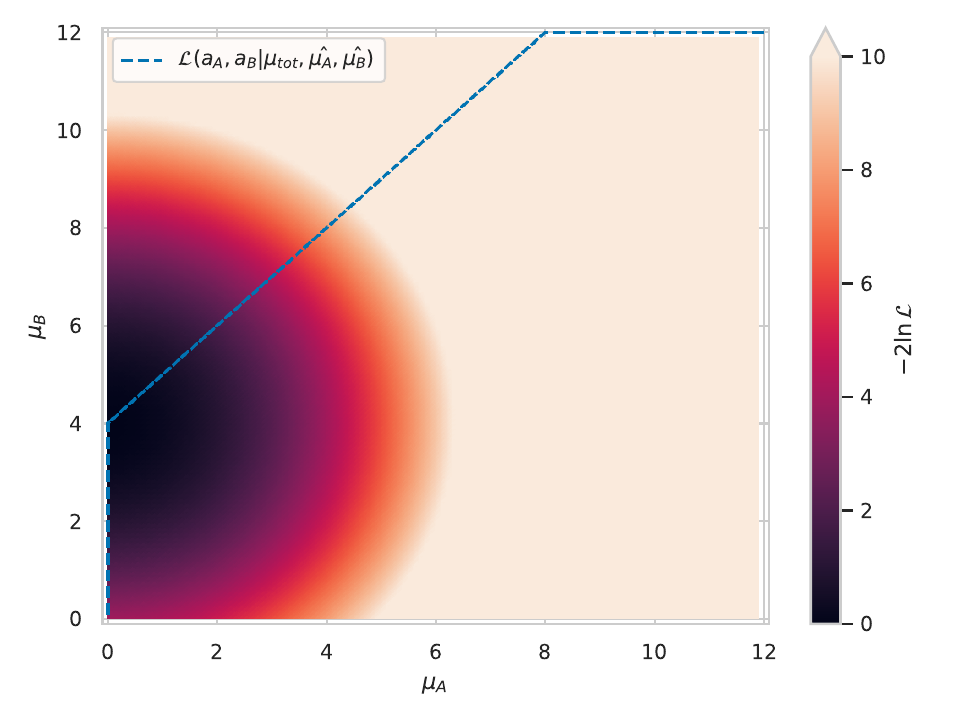}
         \includegraphics[width=\linewidth]{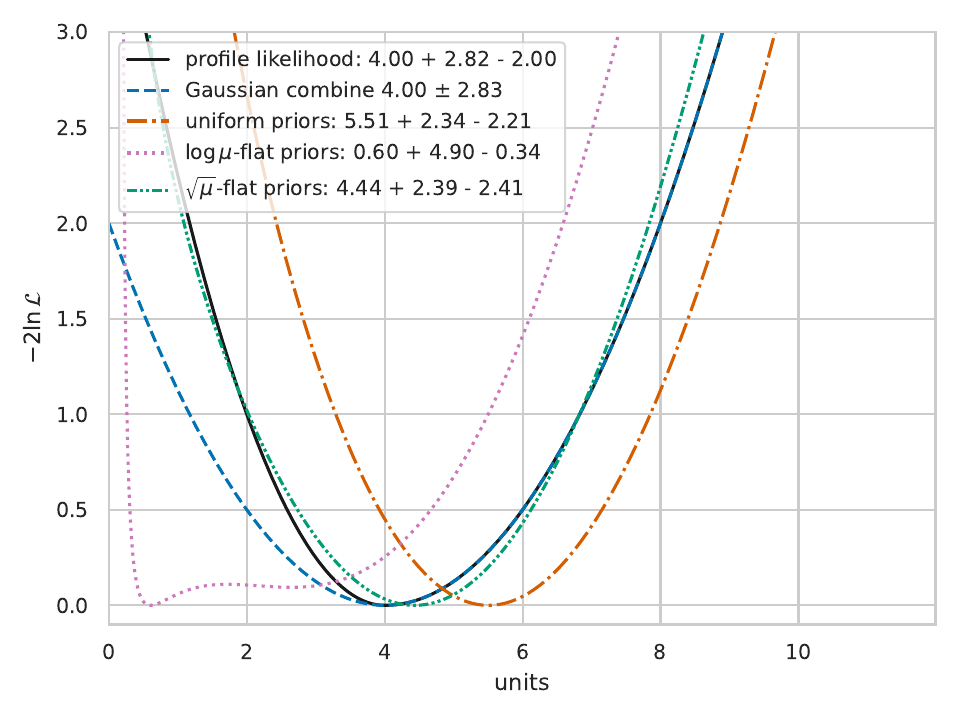}
    
    \caption{Comparison of profile likelihood and posterior constructions for an example with $a_A = 0 \pm 2$ and $a_B = 4 \pm 2$. In this particular example, the 1/$\sqrt\mu$ prior gives close agreement with the frequentist construction while other priors result in different distributions.}
    \label{fig:profileexample1}
\end{figure}

The data is not sensitive to $\mu_A$ or $\mu_B$ due to how we simulate the component group. Therefore, in a frequentist perspective, we can profile over $\mu_A$ and $\mu_B$. In a Bayesian framework, we can marginalize over $\mu_A$ and $\mu_B$ instead with some prior. We also show in Figures \ref{fig:profileexampleboring} and \ref{fig:profileexample1} a usual Gaussian combination of the two results where the means are summed and uncertainties added in quadrature.

We first consider the frequentist framework. In this view, the data has no sensitivity to $\mu_A$ or $\mu_B$ (only to $\mu_{tot}$) and we do not care to estimate the uncertainties on these parameters. We therefore treat these as nuisance parameters and profile them out to find $\mathcal{L}(a_A, a_B | \mu_{\mathrm{tot}}, \hat{\mu_A}, \hat{\mu_B})$. That is, at every $\mu_{tot}$, we find the most likely combination of $\mu_A$ and $\mu_B$ that give this $\mu_{tot}$ and take the likelihood there as the likelihood on that $\mu_{tot}$. The plots in Figures \ref{fig:profileexampleboring} and \ref{fig:profileexample1} show the resulting profile likelihood on the total activity. Note that the most likely total activity occurs at the direct sum of the most likely activities of $A$ and $B$, which makes sense in the frequentist framework. 

In a Bayesian framework, we must choose a prior for the activities, which should in principle be independently chosen and applied to each of the component activities. However, the choice of prior strongly impacts the posterior distribution for the total activity when the likelihood of one of the component activities occurs near the physical boundary of the parameter space. For the estimate of the mean of a Gaussian, the Jeffreys prior, an uninformative prior, is the uniform distribution \cite{jeffreys}. The uniform prior coincides with the frequentist treatment when the likelihoods occur on the interior of the parameter space, as expected, and as occurs in Figure \ref{fig:profileexampleboring}. However, for the case in Figure \ref{fig:profileexample1}, the uniform prior over-predicts the most likely value of the total activity compared to the frequentist treatment and naive expectation. Two other common choices for priors are also shown, a prior flat in $\log\mu$ and a prior flat in $\sqrt{\mu}$, which also yield different results between the interior and near the border of the parameter space. In the examples shown here, these priors are supported on the $\mu = [0.1, 15]$ interval, the choice of which strongly impacts the posterior given that these priors diverge at zero.

There are a few comparisons to make between the treatments. Clearly, the peaks of the likelihoods and Bayesian posteriors occur in slightly different places, depending on the priors chosen. In our view, this is a drawback of the  necessity to choose a prior under a Bayesian treatment. A uniform prior matches the frequentist prescription well on the interior of the parameter space, but at the border, there is a strong bias introduced by the choice of prior. It is not possible to recommend a single prior distribution to account for cases on both the interior and edge of the parameter space. Alternatively, one could try to construct priors that result in a posterior distribution identical to the profile likelihood shown in Figure \ref{fig:profileexample1}. We do not analytically specify this choice of prior, which is not unique, but the posterior can be trivially computed without a complete specification of the prior itself.

The frequentist treatment matches well what one might naively expect from the simple Gaussian combination, with the most likely total activity as the direct sum of the two assays. The upper side of the profile likelihood corresponds exactly with the Gaussian approximation, while the lower side follows the curve of $\mu_B$. Since, for fits to data, we are often not concerned with giving additional freedom to the fit to assign less activity, this suggests that the Gaussian approximation works quite well even in this case.

\begin{figure}[h! tb]
    \centering
     \includegraphics[width=\linewidth]{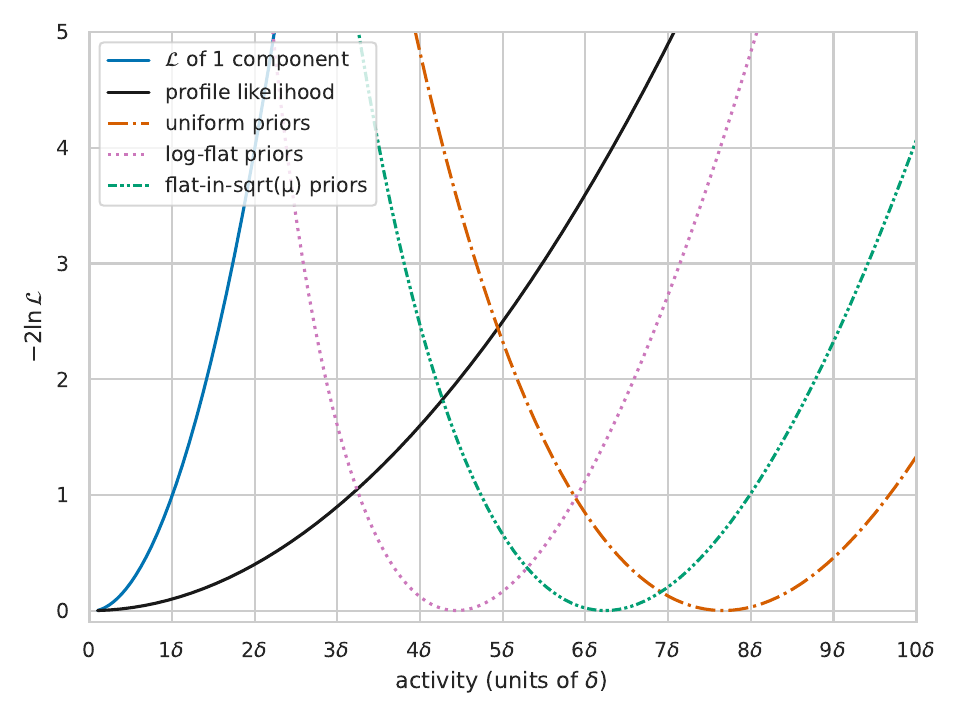}
    \caption{Combination of 10 components with an identical assay result at $<10$ units at 90\% CL, showing the profile likelihood and posterior under several different choices of priors. }
    \label{fig:upperlimitsexample}
\end{figure}

As an extreme example, let us take a case where we combine 10 components into one component group, each of which has an identical assay result at $<1.65 \delta$ units at 90\% CL (corresponding to a measurement of $0 \pm \delta$ in our interpretation from section \ref{sec:basics}). In Figure \ref{fig:upperlimitsexample}, we show the profile likelihood over these ten components, as well as the posterior distribution from an assumption of independent uniform priors on the ten components. As can be seen, these results are quite different, with the central value for the total activity at two very different activities. Here, the profile likelihood result makes the most sense: it should be the case that the most likely total activity is the sum of the most likely component activities since the components and assays are all independent. The posterior distribution instead places the most probable activity at a value corresponding to a $>2 \sigma$ fluctuation away from all the assay results, if we compare to the profile likelihood curve. As this is not a likely result, it does not seem reasonable that this should be the most probable value, suggesting that this choice of prior (independent uniform priors of each of the components) is not reasonable. Other choices of priors lead to a similar ``over-prediction" of total activity compared to the naive expectation. In the case of the $\log\mu$-flat and $\sqrt{\mu}$-flat priors, a different interval extending further towards zero would pull the posterior to lower activities. The choice of prior and interval could correspond to an experimenter's degree of belief about the radioactivity of a particular sample compared to the limit of detection, for example. 

Based on these examples, we find that combining sub-component activities using the simple Gaussian approach is a reasonable approach. It gives an intuitive result for the best estimate and is approximately equal to the profile likelihood method at higher activity levels most relevant in background modeling fits. Although it does not penalize fits to lower activity as much as the true profile likelihood, this is an acceptable compromise for the simplicity of construction. For a Bayesian treatment, the choice of prior has a strong impact on the resulting posterior, and it does not seem possible to recommend a choice of prior applicable to all situations. The Bayesian treatment can be made to trivially coincide with the frequentist treatment through a particular choice. A more informed choice of prior and likelihood can be made with additional assay information, such as the complete likelihood, as opposed to only a limit of detection report.

\section{Conclusions}

We have proposed a new method for averaging the results of radioassays to account for part-to-part radiopurity variation among samples of otherwise identical parts. We have performed a series of Monte Carlo pseudoexperiments demonstrating that the proposed additive-variance method generally outperforms other proposed treatments under a range of different realistic scenarios. Under several scenarios, coverage is not ideal for the additive-variance treatment, but is routinely better than the alternative treatments and is considerably simpler than applying a full Neyman construction to guarantee coverage. The additive-variance method is suitable for most radioassay techniques. However, given that gamma assays are not generally performed multiple times on the same parts due to their time requirements, the method is probably most useful for ICP-MS or other methods repeatable in a shorter time frame. Additionally, we caution that averaging should not be performed on samples that are obviously different or subject to different handling; the experimenter must still use their best judgment for their specific application. We further recommend that multiple assays are conducted for individual parts when possible to evaluate the part-to-part variation, particularly when it is not feasible to non-destructively assay the parts actually entering the experiment.

Finally, we note that the combination of component activities may follow a simple Gaussian approximation to a generally acceptable accuracy. We note that the choice of prior in a Bayesian treatment has important impact on the posterior, particularly when activities are close to the boundary of the parameter space, i.e. close to zero activity. Additional information from reported assays, such as the likelihood and details of the statistical treatment, can improve the choice of prior by providing additional context for a reported result.

\section{Acknowledgments}

We thank the LEGEND collaboration for its inspiration for this work and for providing an opportunity to apply the proposed methods. We thank Ruben Saakyan for useful comments. This work was supported by the US Department of Energy under grant number DE-FG02-97ER41020.

\nocite{*}

\bibliography{apssamp}

\end{document}